\documentclass[twocolumn, showpacs, prl,superscriptaddress]{revtex4}

\usepackage{graphicx}

\usepackage{dcolumn}

\usepackage{color}
\usepackage{bm}

\begin{document}

\title{Spin Susceptibility of Interacting Two-dimensional Electrons with Anisotropic Effective Mass}

\date{\today}

\author{T.\ Gokmen}
\affiliation{Department of Electrical Engineering, Princeton
University, Princeton, NJ 08544}

\author{Medini\ Padmanabhan}
\affiliation{Department of Electrical Engineering, Princeton
University, Princeton, NJ 08544}

\author{E.\ Tutuc}
\altaffiliation{Present address: Microelectronics Research Center,
University of Texas at Austin, Austin, TX 78758, USA}
\affiliation{Department of Electrical Engineering, Princeton
University, Princeton, NJ 08544}

\author{M.\ Shayegan}
\affiliation{Department of Electrical Engineering, Princeton
University, Princeton, NJ 08544}

\author{S.\ De Palo}
\affiliation{INFM-CNR DEMOCRITOS National Simulation Center, Trieste,
Italy} \affiliation{Dipartimento di Fisica Teorica,
Universit$\grave{a}$ di Trieste, Strada Costiera 11, 34014
Trieste, Italy}

\author{S.\ Moroni}
\affiliation{INFM-CNR DEMOCRITOS National Simulation Center, Trieste,
Italy}
\affiliation{International School for Advanced Studies (SISSA), I-34014Trieste, Italy}

\author{Gaetano Senatore}
\affiliation{INFM-CNR DEMOCRITOS National Simulation Center, Trieste,
Italy} \affiliation{Dipartimento di Fisica Teorica,
Universit$\grave{a}$ di Trieste, Strada Costiera 11, 34014
Trieste, Italy}

\begin{abstract}

We report measurements of the spin susceptibility in dilute ({\it
r}$_{s}$ up to $\approx$ 10) AlAs two-dimensional (2D) electrons
occupying a single conduction-band valley with an anisotropic
in-plane Fermi contour, characterized by longitudinal and
transverse effective masses, {\it m}$_{l}$ and {\it m}$_{t}$. As
the density is decreased, the spin susceptibility is significantly
enhanced over its band value, reflecting the role of interaction.
Yet the enhancement is suppressed compared to the results of
quantum Monte Carlo based calculations that take the finite
thickness of the electron layer into account but assume an
isotropic effective mass equal to $\sqrt{m_{l}.m_{t}}$. Proper
treatment of an interacting 2D system with an anisotropic
effective mass therefore remains a theoretical challenge.

\end{abstract}


\maketitle

The low-temperature electronic properties of a clean,
two-dimensional electron system (2DES) are dominated by
electron-electron interaction at low enough densities where the
Coulomb energy is much larger than the kinetic (Fermi) energy. In
particular, the spin susceptibility of a dilute 2DES is expected to
increase significantly over its band value as the density is lowered
\cite{CeperleyPRB1989, AttaccalitePRL2002}. An increase of the
susceptibility has indeed been observed recently in several 2DESs
\cite{OkamotoPRL1999, VitkalovPRL2001, ShashkinPRL2001,
PudalovPRL2002, ZhuPRL2003, PrusPRB2003, TutucPRB2003,
VakiliPRL2004, ShkolnikovPRL2004, TanPRB2006}. The observed
enhancements are qualitatively explained by calculations for an
ideal 2DES although for a quantitative agreement, the properties of
a real 2DES have to be taken into account \cite{DePaloPRL2005,
ZhangPRB2005}. Specifically, when the electrons occupy a $single$
conduction-band valley with an $isotropic$ in-plane Fermi contour,
such as the 2DESs in either a GaAs/AlGaAs heterostructure
\cite{ZhuPRL2003}, or in a narrow AlAs quantum well
\cite{VakiliPRL2004}, the quantum Monte Carlo (QMC) calculations
quite accurately describe the experimental data, once the finite
thickness of the 2DES is included \cite{DePaloPRL2005}.

Here we report measurements of the spin susceptibility for 2DESs
confined to AlAs quantum wells where, thanks to the application of
symmetry-breaking in-plane strain, the electrons occupy a single
conduction-band valley with an $anisotropic$ Fermi contour,
characterized by longitudinal and transverse effective masses, $m_l$
and $m_t$. We compare the measured susceptibility values with the
results of QMC based calculations that take the finite layer
thickness into account but resort to a {\it simple mapping} of the
anisotropic system onto an isotropic one with an effective mass,
{\it m}$_{b}=\sqrt{m_{l}.m_{t}}$ \cite{note1}. As usual, the
electronic coupling parameter is defined in terms of the electron
density ($n$) as the mean inter-electron separation measured in
units of the effective Bohr radius $r_s = 1/\sqrt{{\pi}n}a_B^{*}$.
Here $a_B^{*} = (\epsilon/m_b)a_B$ where $\epsilon$ is the AlAs
dielectric constant, $m_b$ is the effective mass in units of the
free electron mass, and $a_B=0.529${\AA}. Note that $r_s$ can be
equivalently written as the ratio of the Coulomb energy to the Fermi
energy of the 2DES. We find that the experimental values of
susceptibility fall well below the calculated values. The results
highlight the need for a more proper treatment of an interacting
electron system with an anisotropic effective mass.

Bulk AlAs has conduction band minima at the six equivalent X
points of the Brillouin zone. The Fermi surface consists of six
half ellipsoids (three full ellipsoids or valleys), each with an
anisotropic mass ($m_{l}$=1.05 and $m_{t}$=0.205, in units of free
electron mass). We denote these valleys as $X$, $Y$, and $Z$
valleys, according to the direction of their major axes, [100],
[010], and [001], respectively. In bulk AlAs these three valleys
are degenerate, but this degeneracy is lifted in quantum well (QW)
structures \cite{ShayeganPSS2006}. Confinement in the growth
([001]) direction, lowers the energy of the $Z$ valley which has a
larger mass along [001] and a smaller, isotropic mass in the
plane. In a narrow AlAs QW with a width smaller than $\approx$
5nm, the electrons indeed occupy this out-of-plane valley. The
spin susceptibility of such a 2DES, with an $isotropic$ in-plane
effective mass was measured by Vakili {\it et al.}
\cite{VakiliPRL2004}, and was found to be in excellent agreement
with the results of QMC calculations \cite{VakiliPRL2004,
DePaloPRL2005}. It is possible, however, to confine the 2D
electrons to the $X$ and $Y$ valleys whose major axes lie in the
plane. This comes about thanks to the lattice mismatch between
GaAs and AlAs: Since GaAs has a smaller lattice constant than
AlAs, an AlAs QW grown on a GaAs substrate feels a compressive,
bi-axial strain in the plane, which favors the occupation of the
two in-plane valleys. Indeed, these valleys are occupied for AlAs
QW widths larger than $\approx$ 5nm. Furthermore, the degeneracy
of the $X$ and $Y$ valleys can be lifted by applying
symmetry-breaking strain in the plane \cite{ShayeganAPL2003,
ShayeganPSS2006, ShkolnikovAPL2004, ShkolnikovPRL2004,
GunawanPRL2006} so that only one valley, with an $anisotropic$
in-plane effective mass, is occupied. This is the system we
studied in our present work.

We performed measurements on 2DESs confined to AlAs QWs of width
11 and 15nm, grown on (001) GaAs substrates via molecular beam
epitaxy. The AlAs QW is flanked by selectively-doped
Al$_{0.4}$Ga$_{0.6}$As barrier layers \cite{DePoortereAPL2002}. We
fabricated L-shaped Hall bar mesas, along the [100] and [010]
orientations and made Ohmic contacts to the 2DES by depositing
AuGeNi layers and alloying in a reducing environment. Metal front
and back gates were also deposited and used, together with
illumination, to control the 2DES density. As a final step, we
thinned our samples down to $\sim$200 microns and glued them on a
piezoelectric actuator \cite{ShayeganPSS2006, ShayeganAPL2003}.
Using the actuator, we were able to transmit sufficiently large
compressive strain in the [100] direction to transfer all the
electrons to the $X$ valley \cite{ShayeganPSS2006,
ShkolnikovAPL2004}. The density range of 2.2 to
6.7$\times$10$^{11}$cm$^{-2}$ achieved in our samples corresponds
to an $r_{s}$ range of 6 to 10.5, using AlAs dielectric constant
of 10 and the band density-of-states effective mass of
$m_b=\sqrt{m_l.m_t}=0.46$. The densities were determined from the
Hall resistance and the minima of the Shubnikov-de Haas
oscillations of the longitudinal resistance ($R_{xx}$). The
mobility in this density range varied from 1.1 to 3.0 m$^{2}$/Vs
for current along [100] and 2.6 to 5.2 m$^{2}$/Vs for current
along the [010] direction. Measurements were done down to $T$ =
0.3K using the lock-in technique and in a cryostat with a tilting
stage, allowing the angle $\theta$, between the sample normal and
the magnetic field to be varied {\it in situ}. We denote the
parallel and perpendicular components of the total magnetic field
({\it B}$_{tot}$) by $B_\|$ and $B_\bot$, respectively. The
orientation of $B_\|$ for the data we present here was along [100]
(see Fig. 1(b)); in separate cooldowns, we verified that, within
the accuracy of our measurements, the spin-susceptibility does not
depend on the orientation of the magnetic field with respect to
the direction of the major axis of the occupied valley.

\begin{figure}
\centering
\includegraphics[scale=0.97]{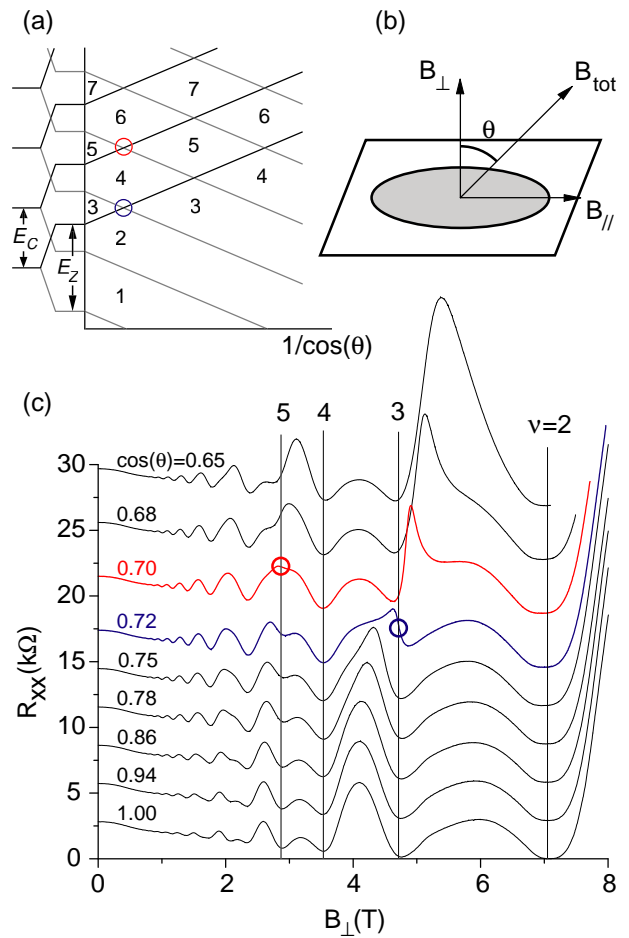}
\caption{(Color online) (a) Schematic energy fan diagram showing
the spin-split Landau levels as a function of tilt angle, $\theta$
. The cyclotron ($E_C$) and Zeeman ($E_Z$) energies and the Landau
level filling factors ($\nu$) are indicated. (b) The in-plane
electron Fermi contour is shown along with the components of the
applied magnetic field. (c) Resistance vs. $B_\bot$ traces at $T$
= 0.3K for 2D electrons ($n$ = 3.4$\times$10$^{11}$cm$^{-2}$)
confined to a 15nm-wide AlAs quantum well. Data are shown for
different values of $cos( \theta)$ as indicated, and the traces
are shifted vertically for clarity. The blue and red traces show
the coincidence for $\nu$ = 3 and $\nu$ = 5. }
\end{figure}

We used the "coincidence" technique \cite{FangPR1968} to measure
the spin susceptibility or, equivalently, $g^{*}m^{*}$, where
$g^{*}$ and $m^{*}$ are the effective Land\'{e}  g-factor and
mass, respectively. Note that $\chi_s = d{\Delta}n/dB \propto
g^{*}m^{*}$, where $\Delta n$ is the net spin imbalance. As shown
in the simple fan diagram in Fig. 1(a), the 2DES energy is
quantized into Landau levels (LLs) which are separated by the
cyclotron energy, $E_C={\hbar}eB_{\bot}/m^{*}$. Each LL is in turn
split into two levels separated by the Zeeman energy,
$E_Z=g^{*}\mu_{B}B_{tot}$. By varying $\theta$, one can control
the ratio between $E_Z$ and $E_C$. When this ratio equals an
integer value at some critical angle, two different LLs coincide
in energy and, if this happens at the Fermi level, then the
resistance becomes a maximum. At the coincidence, the ratio of
Zeeman to cyclotron energy can be written as
$E_Z/E_C=i=g^{*}m^{*}/2cos({\theta}_{i})$, where $i$ is the
difference in the LL index of the crossing levels. Note that if
$g^{*}m^{*}$ does not depend on the spin polarization
$\zeta=\Delta n/n$ then, as seen in the simple fan diagram of Fig.
1(a), coincidences for all the odd (or even) fillings would happen
at the same $\theta$. As we illustrate below, this is not
necessarily the case for our data; we therefore use a coincidence
condition that keeps track of the spin polarization,
$(g^{*}m^{*})_{\nu}/2cos({\theta}_{i,{\nu}})=i$, where $\zeta = i
/ \nu$ is the ratio of the difference between the indices of the
crossing LLs (i.e., the number of the filled, majority spin
levels) and the number of filled LLs. Using the above equation
$g^{*}m^{*}$ can be deduced at different spin polarizations by
measuring the coincidence angle at different fillings,
$\theta_{i,{\nu}}$.

 In Fig. 1(c) we show characteristic magnetoresistance traces at
different tilt angles. Strong minima are observed at $\theta$ = 0
(bottom trace) at several filling factors, including $\nu$ = 2, 3,
4, and 5. As the sample is tilted, the $R_{xx}$ minimum at $\nu$ =
3 turns into a maximum at $cos({\theta)}= 0.72$ (blue trace)
indicating a coincidence for $\nu$ = 3 at this angle. At a slightly
larger angle ($cos({\theta)}=0.70$, red trace) the $\nu$ = 5 state
goes through a coincidence. The states at even $\nu > 2$ go
through coincidences at higher $\theta$ (data not shown). But note
that the $\nu$ = 2 state remains strong at all tilt angles. This
is because at this density, $E_{Z}$ is larger than $E_{C}$ even at
zero tilt due to the enhanced $g^{*}m^{*}$, as shown in the fan
diagram of Fig. 1(a).

We summarize the measured values of $g^{*}m^{*}$, normalized to the
band values $g_b=2$ and $m_b=0.46$, as a function of $n$ in Fig. 2.
Three aspects of the data are noteworthy. First, the data exhibit a
strong enhancement of $g^{*}m^{*}$ with decreasing $n$,
qualitatively consistent with previous reports for various 2DESs
\cite{OkamotoPRL1999, VitkalovPRL2001, ShashkinPRL2001,
PudalovPRL2002, ZhuPRL2003, PrusPRB2003, TutucPRB2003,
VakiliPRL2004, ShkolnikovPRL2004, TanPRB2006}. Second, at a given
$n$, there is a small but noticeable dependence of $g^{*}m^{*}$ on
the degree of spin polarization ($\zeta$). This is evident from the
data of Fig. 1(c): the $\nu$ = 3 state ($\zeta$ = 0.67) goes through
a coincidence at a slightly smaller $\theta$ than the $\nu$ = 5
state ($\zeta$ = 0.40), reflecting a slightly larger $g^{*}m^{*}$
when $\zeta$ is larger \cite{BparComment}. This dependence, which is
explicitly shown in Fig. 2 and its inset, is consistent with what is
theoretically expected, as we discuss below. Third, experimental
data taken on an 11nm-wide AlAs QW with only one in-plane valley
occupied \cite{ShkolnikovPRL2004} show $g^{*}m^{*}$ values that are
about 10\% to 15\% larger than those for the 15nm QW data shown in
Fig. 2. Clearly, the larger the thickness of the 2DES, the smaller
is the value of the spin susceptibility at a given $n$.

\begin{figure}
\centering
\includegraphics[scale=0.97]{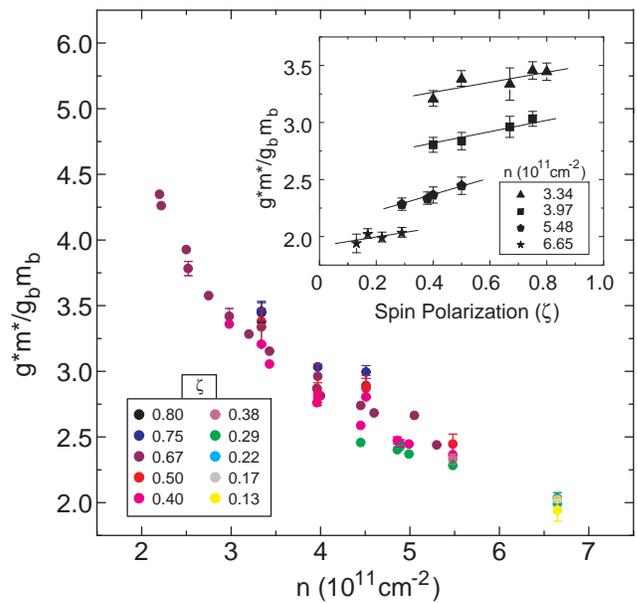}
\caption{(Color online) Spin susceptibility, normalized to band
value, vs. density for 2D electrons confined to a 15nm-wide AlAs
quantum well. Different colors corresponds to different spin
polarizations, $\zeta$. Inset shows the dependence of the spin
susceptibility on $\zeta$ at four different $n$. The error bars
are from the resolution and accuracy of $\theta$ at which a
coincidence occurs, and the lines in the inset are guides to the
eye. }
\end{figure}

For a quantitative understanding of the data, we performed QMC
based calculations of the spin susceptibility for our 2DES
resorting to the simple mapping discussed above and in footnote
\cite{note1}, thus exploiting the scheme for isotropic systems
described in Ref. \cite{DePaloPRL2005}. The results of these
calculations, are shown in Fig. 3 both for a 2DES confined to a
15nm-wide QW ($w$ = 15nm, lower two curves) and for an ideal 2DES
with zero layer thickness ($w$ = 0, upper two curves). For each
case, the QMC predictions are given for the spin susceptibility
determined in the limit of zero spin polarization ($\zeta=0$,
solid curves) and full polarization ($\zeta=1$, dashed curves)
\cite{note2}. For a given layer thickness, the $\zeta$ = 1 curve
is above the $\zeta$ = 0 curve.  This indicates an increasing
$g^{*}m^{*}$ with increasing $\zeta$ which indeed qualitatively
agrees with the experimental data (Fig.  2 inset). Note also that
the finite thickness of the electron layer softens the
electron-electron interaction and suppresses the spin
susceptibility, again in qualitative agreement with the
experimental data, as described in the previous paragraph
($g^{*}m^{*}$ is smaller for the 15nm QW compared to the 11nm QW).
Indeed, it was shown in Ref. \cite{DePaloPRL2005} that once the
finite layer thickness of the 2DES is taken into account, the QMC
predictions $quantitavely$ reproduce the experimental data for
2DESs with an $isotropic$ (in-plane) effective mass. This was
illustrated for the case of AlAs 2D electrons confined to a
4.5nm-wide QW (and occupying the $Z$ valley which has an isotropic
in-plane Fermi contour) and also GaAs 2D electrons at a
GaAs/AlGaAs interface (occupying a single, isotropic valley at the
$\Gamma$-point of the Brillouin zone).

The main point of our paper is that the results of similar
calculations (lower curves in Fig. 3), done for our 2DES, do $not$
accurately describe the experimental data points, but rather
overestimate the spin susceptibility by as much as about 45\%. This
is also the case for the data from the 11nm-wide QW. We believe that
the culprit is the $anisotropic$ effective mass of the 2DES. Note
that for the QMC calculations whose results are shown in Fig. 3, we
mapped  the system with an anisotropic mass onto an effective system
with an isotropic mass  equal to $\sqrt{m_{l}.m_{t}}$. Evidently,
this assumption leads to an overestimation of the spin
susceptibility. We emphasize that two effects that are not included
in the calculations, namely the presence of disorder and parallel
magnetic field \cite{BparComment}, both lead to a further
$enhancement$ of the susceptibility: Slight disorder is expected to
increase the susceptibility \cite{DePaloPRL2005}, and the parallel
field enhances the effective mass, also leading to a larger
susceptibility \cite{TutucPRB2003,BparComment}. The only other
effect that is not fully included in the calculations is the
anisotropy of the effective mass \cite{note3}. Properly
incorporating electron-electron interaction in a 2DES with an
anisotropic effective mass therefore remains a theoretical
challenge.

\begin{figure}
\centering
\includegraphics[scale=0.97]{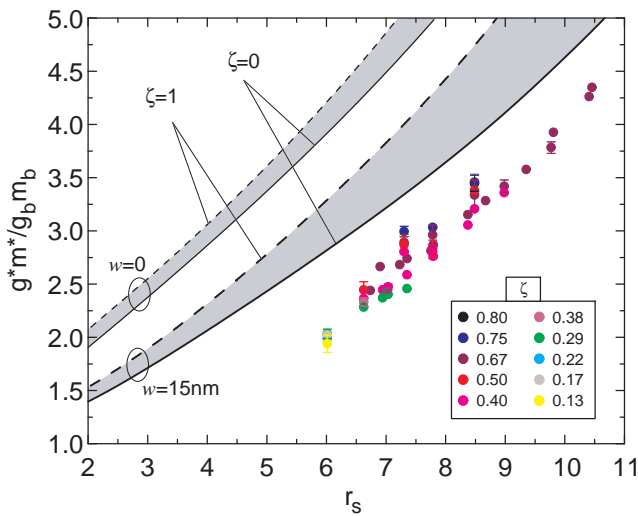}
\caption{(Color online) Spin susceptibility vs. $r_{s}$ plots for both
  experimental data and quantum Monte Carlo calculations. Data points
  are shown as circles and different colors correspond to different
  spin polarizations, $\zeta$. Thin full and dashed lines are results
  of the QMC calculations \cite{DePaloPRL2005} for an ideal 2D electron
  layer (thickness $w=0$), at zero ($\zeta$=0) and full ($\zeta$=1)
  spin polarizations, respectively. Thick full and dashed lines are
  QMC based results for electrons in a 15nm-wide AlAs
  quantum well at zero and full spin polarizations respectively.}
\end{figure}

We thank the NSF for support.

\break

\end{document}